\documentclass[12pt]{article}
\usepackage{amssymb}
\usepackage{amsmath}
\usepackage{hyperref}
\usepackage{authblk}
\usepackage{graphicx}
\usepackage{subfig} 
\begin{document}
\title{\bf Critical behavior of AdS Gauss-Bonnet massive black holes in the presence of external string cloud }

\author[1]{Hadi Ranjbari\thanks{Corresponding author:Email:fhranji@gmail.com}}
\author[2]{Mehdi Sadeghi\thanks{ Email:mehdi.sadeghi@abru.ac.ir}}
\author[3]{M.Ghanaatian\thanks{Email:m\_ghanaatian@pnu.ac.ir}}
\author[4]{Gh.Forozani\thanks{ Email:forozani@pnu.ac.ir}}

\affil[1,3,4]{\small{\em{Department of  Physics, Payame Noor University (PNU), P.O.Box 19395-3697 Tehran, Iran}}}
\affil[2]{\small{\em{Department of Physics, School of  Sciences, Ayatollah Boroujerdi University, Boroujerd, Iran}}}

\date{\today}
\maketitle

\abstract{Following previous study about AdS-Schwarzschild black holes minimally coupled to a cloud of strings in the context of massive gravity \cite{Ghanaatian:2019} and inspired by strong connection between Gauss-Bonnet Gravity and heterotic string theory, in this paper, we first take into account the Gauss-Bonnet term and we study thermodynamics and critical behavior of these black holes in the extended phase space. The effects of Gauss-Bonnet, massive, and string cloud  parameters on the criticality of these black holes has been investigated. It can be seen that the Gauss-Bonnet and massive parameters have opposite effects on the criticality and phase transition of the solutions. We also observe that the increase in the value of the string cloud parameter above a critical value, eliminates the van der Waals like behavior of these solutions. Also, the Joule-Thomson effect is not observed. Then we examine thermal stability of these black holes in canonical ensemble by calculating the heat capacity. In addition, we explore critical behavior in extended phase space by employing heat capacity and consequently, we observe that the results are in agreement with the previous results from the usual method in section 3.}


\section{Introduction} \label{intro}
After the introduction of Einstein's theory of general relativity, at first a series of mathematical singularities appeared as solutions to Einstein's general equation of relativity. The first settled solution of general relativity as a black hole was found by Schwarzschild in 1916. For a long time, black holes were considered to be a mathematical peculiarity, until in 1973 Bekenstein introduced black holes as  interesting thermodynamic systems that follow the laws of usual thermodynamics \cite{Bekenstein:1973}. This progress made us one step closer to a better understanding of quantum gravity \cite{Bardeen:1973}.\\
On the other hand, Einstein's equation generally described an expanding universe, and Einstein first added a cosmological constant to his equation to describe a static universe. But after discovering the expansion of the universe by Hubble in 1931, Einstein ignored the cosmological constant, until 1998, it was discovered that the expansion of the universe is accelerating, imposing a positive value for the cosmological constant. If the cosmological constant is positive, the associated negative pressure will derive an accelerated expansion of the universe, as observed from the Planck Collaboration \cite{Aghanim:2018eyx}. So far, cosmological constant is considered as a constant parameter, but on the contrary, in black hole thermodynamics is regarded as thermodynamic pressure which can vary. From this perspective, an extended phase space appears with a new dimension added, and the negative cosmological constant determines a positive varying thermodynamic pressure in this new framework 
\begin{equation}
P =  - \frac{\Lambda }{{8\pi }},
\end{equation}
to review a few examples of work in the context of extended phase space, see \cite{Dayyani:2018fuz}-\cite{Hendi:2015pda}.\\
Among the modified gravity theories, one of the interesting proposals  is Lovelock theory, which is reduced to Einstein's theory in three and four dimensions, but in five dimensions and more, it includes higher curvature terms. The first additional term is Gauss-Bonnet (GB) gravity where is quadratic in curvature tensors and  leads to field equations that are second-order unto metric derivatives and do not engage in ghosts \cite{Boulware:1985},\cite{Cai:2002}. It was also shown that the GB theory can be derived from the low-energy limit of heterotic string theory \cite{Zwiebach:1985},\cite{Gross:1986}. One of the promising theories in modern theoretical cosmology is the scalar-Einstein-Gauss-Bonnet gravity theory \cite{Nojiri:2005vv},\cite{Nojiri:2006je} which is motivated  by string theory and shows how the string theory affects the primordial acceleration of the universe. Many other aspects of GB theory have been studied in literature \cite{Kleihaus:2011tg}-\cite{Li:2013}.\\
One of the most consistent theory of gravity that modifies GR by endowing the graviton with a nonzero mass, is dRGT massive gravity \cite{de Rham:2010kj}. This theory is ghost free and avoids discontinuities in the limit where the graviton mass goes to zero. Also, the theory of general relativity is modified by massive gravity at large distances, explains the accelerated expansion of the universe, without resorting to the concept of dark energy. In addition , the cosmological solutions of massive gravity and its expanded types, such as bimetric gravity \cite{Hassan:2012zd}, can reflect  late-time acceleration compatibility with observations \cite{D'Amico:2011eto}-\cite{Akrami:2015qga}.\\
However, string theory predicts the existence of a graviton, but we do not have a successful quantum theory of gravity. Superstring theory is a theory that considers the particles and fundamental forces of nature as vibrations of tiny supersymmetric strings. The idea of taking fundamental particles as vibration modes of one-dimensional string objects plays an essential role in these theories. The fast accelerated expansion of the universe in the period of inflation can be attributed to the stretch of such cosmic strings that penetrated everywhere in our observed universe \cite{Henry Tye:2008uv}. A cloud of strings, an aggregation of one dimensional objects in a certain geometrical frame, was proposed by Letelier \cite{Letelier:1979}. The gravitational effects of matter in the form of sting cloud are studied \cite{Letelier:1979}-\cite{Ghosh:2014dqa}.\\
On the other side, when a control parameter such as temperature is changed in a thermodynamic system, the system may change to a different macroscopic state that is more stable, which this mutation is called phase transition. In a phase transition, a thermodynamic potential such as free energy becomes non-analytic. Although the AdS black holes in the radiation can remain stable in the heat stability, but at a certain critical temperature, there is a phase transition called Hawking-Page phase transition \cite{Hawking:1983}. This phase transition is mainly observed in Einstein's general relativity family such as Gauss-Bonnet gravity \cite{Wang:2019vgz} and other theories like dilaton gravity \cite{Cai:2007wz}, magnetic black brane \cite{Huang:2010qz}, black Dp-branes and R-charged black holes with an IR Cutoff \cite{Cai:2007vv} and BTZ black hole \cite{Eune:2013qs}. There is also a resemblance between ``small black hole/large black hole"(SBH/LBH) phase transition and the liquid-gas phase transition, that this critical behavior is called Van der Waals-like behavior. It can be seen that this behavior is commonly found in charged AdS black holes and in massive gravity theories, as well as the coupled theory with a cloud of strings. \cite{Chamblin:1999tk}-\cite{Kubiznak:2012wp}.\\
Recently, the effects of a cloud of strings on the extended phase space of Einstein-Gauss-Bonnet AdS black hole is studied and the Van der Waals-like behavior in absence of the GB term is observed \cite{Ghaffarnejad:2018tpr}. Also, in the previous work \cite{Ghanaatian:2019}, we investigated the effects of the external string cloud on the Van der Waals like behavior of AdS-Schwarzschild black holes in massive gravity. With these explanation, in this paper, we study critical behavior of AdS Gauss-Bonnet massive black holes in the presence of external string cloud. One of the motivations of this study is to investigate the simultaneous effects of the massive, Gauss-Bonnet and cloud of strings terms on critical points and other is study the effect of each of these parameters on critical behavior by keeping the other two parameters in constant.\\
This paper is organized as follows. In section 2, the solutions of AdS-GB black holes in dRGT massive gravity minimally coupled to a cloud of strings are introduced and the metric function and its diagrams in different modes are investigated. In section 3, we examine the first law of  thermodynamics of these black holes and employ extended phase space thermodynamics to explore critical points. Also, we study the behavior of system along the coexistence line by plotting isothermal curves in $P - T$ diagrams. We then investigate the possibility of the Joule-Thomson effect in our model by drawing isenthalpic curves in $T - P$ plan. In the following, the critical exponents are calculated  in our model. The thermal stability of the solutions in canonical ensemble are studied in section 4, and more, a search for critical behavior in extended phase space has been performed using heat capacity. Finally, in last section we will present our conclusions.  

\section{Black hole in GB-massive gravity minimally coupled to a cloud of strings }
 \label{sec2}
Let us start with an AdS GB-massive gravity in 5-dimensions in the presence of external string cloud. The action is as follows,
\begin{equation}
I =  - \frac{1}{{16\pi }}\int {{d^5}x\sqrt { - g} [R - 2\Lambda  + {\lambda _{gb}}{{\cal L}_{gb}} + {m^2}\sum\limits_{i = 1}^4 {{c_i}{{\cal U}_i}(g,f)} ]}  + \int_\Sigma  {{{\rm N}_{\rm{P}}}} \sqrt { - \chi } d{\lambda ^0}d{\lambda ^1},
\end{equation}
where $ R $ is the scalar curvature, $\Lambda  = {{ - 6} \over {{L^2}}}$ is cosmological constant with $ L$ as the cosmological constant scale, $f$ is a fixed rank-2 symmetric tensor known as reference metric and $m$ is the massive parameter. The last part called a Nambu-Goto action, in which ${{\rm N}_{\rm P}}$ is a positive quantity and is related to the tension of string, $({\lambda ^0},{\lambda ^1})$ is a parametrization of the world sheet $\Sigma $ and $ \chi $ is the determinant of the induced metric \cite{Ghosh:2014pga}-\cite{Graca:2016cbd}
\begin{equation}
{\chi _{ab}} = {g_{\mu \nu }}\frac{{\partial {x^\mu }}}{{\partial {\lambda ^a}}}\frac{{\partial {x^\nu }}}{{\partial {\lambda ^b}}}.
\end{equation}
$\mathcal{L}_{gb}$ is the Gauss-Bonnet term of gravity with ${\lambda _{gb}}$ its dimensionless coupling. $\mathcal{L}_{gb}$ is given by
\begin{equation}
\mathcal{L}_{gb}=R^2-4R_{\mu \nu}R^{\mu \nu}+R_{\mu \nu \rho \sigma }R^{\mu \nu \rho \sigma},
\end{equation}
where ${R_{\mu \nu }}$ and ${R_{\mu \nu \rho \sigma }}$ are Ricci and Riemann tensors respectively.\\
In massive term, $ c_i $'s are constants and $ \mathcal{U}_i $'s are symmetric polynomials of the eigenvalues of the $ 5\times5 $ matrix $ \mathcal{K}^{\mu}_{\nu}=\sqrt{g^{\mu \alpha}f_{\alpha \nu}} $
 \begin{align}\label{7}
  & \mathcal{U}_1=[\mathcal{K}],\nonumber\\
  & \mathcal{U}_2=[\mathcal{K}]^2-[\mathcal{K}^2],\nonumber\\
  &\mathcal{U}_3=[\mathcal{K}]^3-3[\mathcal{K}][\mathcal{K}^2]+2[\mathcal{K}^3],\nonumber\\
  & \mathcal{U}_4=[\mathcal{K}]^4-6[\mathcal{K}^2][\mathcal{K}]^2+8[\mathcal{K}^3][\mathcal{K}]+3[\mathcal{K}^2]^2-6[\mathcal{K}^4],
 \end{align}
the square root in $ \mathcal{K} $ means $(\sqrt {\rm A} )_\nu ^\mu (\sqrt {\rm A} )_\lambda ^\nu  = {\rm A}_\lambda ^\mu $ and the rectangular brackets denote traces. \\
A generalized version of $ f_{\mu \nu} $ was proposed in \cite{Sadeghi:2019}, \cite{Sadeghi:2018vrf} with the form
$ f_{\mu \nu} = diag(0,0,c_0^2h_{ij})$, where ${h_{ij}} = {1 \over {{L^2}}}{\delta _{ij}}$.
The values of $ \mathcal{U}_i $'s are calculated as below,
\begin{align}\label{8}
  & \mathcal{U}_1=\frac{3c_{0}}{r}, \,\,\,  \,\,\, \mathcal{U}_2=\frac{6c_0^2}{r^2},\,\,\,\,\mathcal{U}_3=\frac{6c_0^3}{r^3},\,\,\,\,\mathcal{U}_4=0.\nonumber
   \end{align}
The energy-momentum tensor for a cloud of strings is given by 
\begin{equation}
{{\rm T}^{\mu \nu }} = \rho {\Sigma ^{\mu \sigma }}\Sigma _\sigma ^\nu /\sqrt { - \chi } ,
\end{equation}
where $\rho $ is the proper density of a string cloud and $\Sigma _\nu ^\mu $ is the spacetime bivector
\begin{equation}
{\Sigma ^{\mu \nu }} = {\varepsilon ^{ab}}\frac{{\partial {x^\mu }}}{{\partial {\lambda ^a}}}\frac{{\partial {x^\nu }}}{{\partial {\lambda ^b}}},
\end{equation}
in which ${\varepsilon ^{ab}}$ is Levi-Civita tensor.\\
Conservation of the energy-momentum tensor, ${\nabla _\nu }{{\rm T}^{\mu \nu }} = 0$ results in,
\begin{equation}
{\partial _\mu }(\sqrt { - g} \rho {\Sigma ^{\mu \sigma }}) = 0.
\end{equation}
The equation of motion is obtianed by variation of the action with respect to the metric tensor ${g_{\mu \nu }}$
\begin{equation}
{G_{\mu \nu }} + \Lambda {g_{\mu \nu }} + {{\rm H}_{\mu \nu }} + {m^2}{X_{\mu \nu }} = {{\rm T}_{\mu \nu }},
\end{equation}
where ${G_{\mu \nu }}$ is the Einstein tensor, ${{\rm H}_{\mu \nu }}$ and ${X_{\mu \nu }}$ are 
\begin{equation}
{{\rm H}_{\mu \nu }} =  - \frac{{{\lambda _{gb}}}}{2}(8{R^{\rho \sigma }}{R_{\mu \rho \nu \sigma }} - 4R_\mu ^{\rho \sigma \lambda }{R_{\nu \rho \sigma \lambda }} - 4R{R_{\mu \nu }} + 8{R_{\mu \lambda }}R_\nu ^\lambda  + {g_{\mu \nu }}{{\cal L}_{gb}}),
\end{equation}
\begin{equation}
\begin{split}
{X_{\mu \nu }} =  - \frac{{{c_1}}}{2}(\mathcal{U}_1{g_{\mu \nu }} - {\mathcal{K} _{\mu \nu }}) - \frac{{{c_2}}}{2}( \mathcal{U}_2{g_{\mu \nu }} - 2 \mathcal{U}_1{\mathcal{K} _{\mu \nu }} + 2\mathcal{K} _{\mu \nu }^2) - \frac{{{c_3}}}{2}( \mathcal{U}_3{g_{\mu \nu }} - 3 \mathcal{U}_2{\mathcal{K} _{\mu \nu }}\\
 + 6 \mathcal{U}_1\mathcal{K} _{\mu \nu }^2 - 6\mathcal{K} _{\mu \nu }^3) - \frac{{{c_4}}}{2}( \mathcal{U}_4{g_{\mu \nu }} - 4 \mathcal{U}_3{\mathcal{K} _{\mu \nu }} + 12 \mathcal{U}_2\mathcal{K} _{\mu \nu }^2 - 24 \mathcal{U}_1\mathcal{K} _{\mu \nu }^3 + 24\mathcal{K} _{\mu \nu }^4).
\end{split}
\end{equation}
We consider the following metric ansatz for a five-dimensional static spherically symmetric metric
\begin{equation}
d{s^2} =  - f(r)d{t^2} + \frac{{d{r^2}}}{{f(r)}} + {r^2}{h_{ij}}d{x^i}d{x^j}.
\end{equation}
Inserting this  ansatz into the Eq (9) yields  two solutions for $f(r)$
\begin{equation}
{f_ \pm }(r) = k + \frac{{{r^2}}}{{4{\lambda _{gb}}}}\left[ {1 \pm \sqrt {1 - 8{\lambda _{gb}}(\frac{1}{{{L^2}}} - \frac{{{b^4}}}{{{r^4}}} - \frac{{2a}}{{3{r^3}}}) - 8{m^2}{\lambda _{gb}}(\frac{{c_0^3{c_3}}}{{{r^3}}} + \frac{{c_0^2{c_2}}}{{{r^2}}} + \frac{{{c_0}{c_1}}}{{2r}}} )} \right],
\end{equation}
in which ${b^4}$ is an integration constant and $a$ is a real positive constant known as string cloud parameter. Since black hole should have an event horizon we pick up ${f_ - }(r)$. To calculate the radius of the event horizon, $ r_0 $, we set ${f_ - }( r_0) = 0$, so we have
\begin{equation}
{b^4} \equiv {m_0} = r_0^4\left[ {\frac{1}{{{L^2}}} - \frac{{2a}}{{3r_0^3}} + \frac{k}{{r_0^2}} + \frac{{2{\lambda _{gb}}{k^2}}}{{r_0^4}} + {m^2}\left( {\frac{{c_0^3{c_3}}}{{r_0^3}} + \frac{{c_0^2{c_2}}}{{r_0^2}} + \frac{{{c_0}{c_1}}}{{2{r_0}}}} \right)} \right],
\end{equation}
where ${m_0}$ is related to the total mass of the black hole with $M = {{3{V_3}{m_0}} \over {16\pi }}$ where ${V_3} = {{4\pi } \over 3}$, is volume of the three dimensional unit sphere as plane or hyperbola.\\
To provide the estimated value for the dimensionless coupling coefficient associated with the GB-term, the well-defined constraint of the vacuum solution (${b^4} = 0$, $a = 0$ and $m = 0$) leads to $0 \le {\lambda _{gb}} \le {{{L^2}} \over 8}$. Besides, the causality and positive requirement of the boundary energy density in holography  requires that $ - 7.72 \le {\lambda _{gb}} \le 9.20$, \cite{Ghaffarnejad:2018tpr}, then we can estimate $L \simeq 8.579044$ for maximum value of the cosmological constant scale in vacuum solution. However we see that if $r \to 0$ then $f(r)$ approaches $k$ so there is no $r = 0$ singularity in vacuum solution. Also in non-vacuum regime we observe that if $r \to 0$ then $f(r)$ approaches $k - \sqrt {{{{m_0}} \over {2\lambda_{gb} }}}$, it means adding the Gauss-Bonnet  term to the action causes the causal singularity to be removed.  \\ 
In order to investigate the effects of cloud string parameter and GB parameter on the metric function of gravitational theory we plot $f(r) - r$ diagrams in the following (Figure 1). We observe that, due to the addition of the GB parameter, singularity is gone. By comparing the diagrams (a) and (b), we observe that in flat topology,  the number of horizons decreases by decreasing the value of the GB parameter. It is also clear from the difference between the two diagrams (c) and (d) that the number of horizons increases with increasing GB parameter. Contrary to this, the comparison of the diagrams (a) and (c) shows that the increase of the string cloud parameter, in contrast to the GB parameter, has a decreasing role in the number of horizons. All in all, it can be said, the existence of a maximum of 3 roots in $f(r) - r$ diagrams is obvious. It should be noted that in diagrams (a) through (d), the change in the parameters of ${\lambda _{gb}}$ and $a$, except for the effect on the number of horizons, does not have a significant effect on the metric function, especially on asymptotical behavior, but for the negative value for ${\lambda _{gb}}$, the diagrams (e) and (f) show that for radii less than a certain value, the metric function has no value and so is not defined. Also, for the negative values of ${\lambda _{gb}}$, the asymptotical behavior is different from that of the positive values of ${\lambda _{gb}}$. 
\begin{figure}[ht]
    \centering
\subfloat[a]{\includegraphics[width=4.5cm]{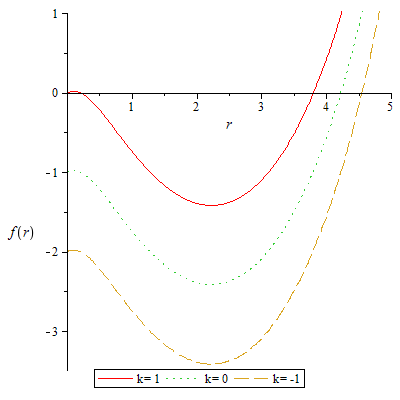}}
\qquad
\subfloat[b]{\includegraphics[width=4.5cm]{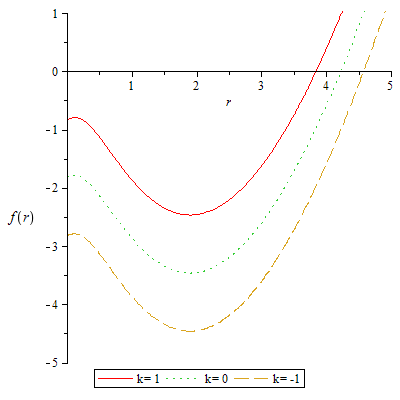}}
\qquad
\subfloat[c]{\includegraphics[width=4.5cm]{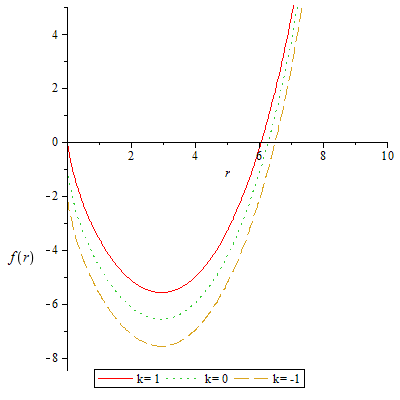}}
\qquad
\subfloat[d]{\includegraphics[width=4.5cm]{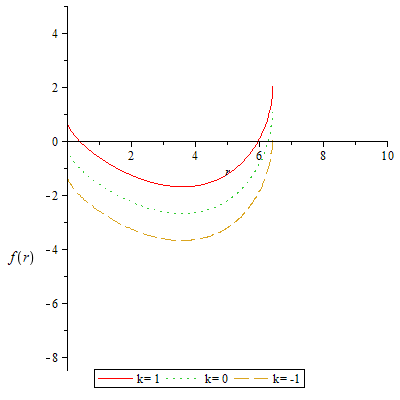}}
\qquad
\subfloat[e]{\includegraphics[width=4.5cm]{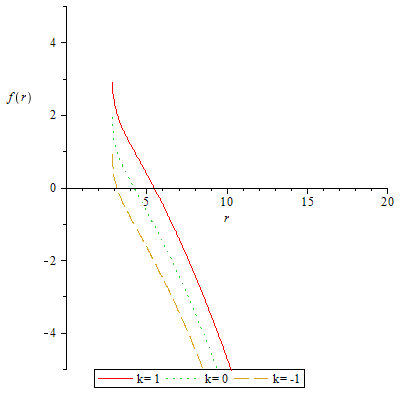}}
\qquad
\subfloat[f]{\includegraphics[width=4.5cm]{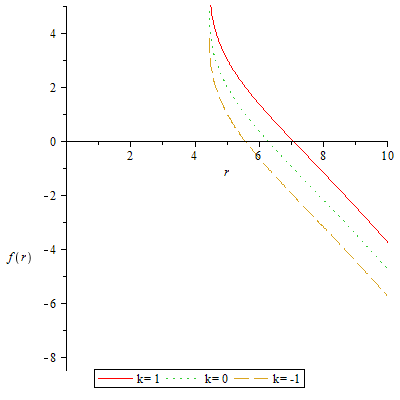}}
\caption{ $f(r)- r$  for $d=5$, ${L^2} = 6$, ${c_0} = 1$, ${c_1} = 0.6$, ${c_2} = -2$, ${c_3} = 0.5$, ${m_0} = 2$, and $m = 2.1$; (a) : $a$=1, ${\lambda _{gb}} =1$, (b): $a$=1, ${\lambda _{gb}} = 0.3$, (c): $a$=60, ${\lambda _{gb}} =1$, (d): $a$=60, ${\lambda _{gb}} =9$,  (e): $a$=1, ${\lambda _{gb}} = -0.3$, and (f): $a$=60, ${\lambda _{gb}} = -0.3$.}
\end{figure}

\section{Thermodynamics and critical behavior in the extended phase space}
\label{sec3}
In this section, we first calculate the thermodynamic quantities of AdS black holes in GB-massive gravity minimally coupled to a cloud of strings in the extended phase space and we present the first law of thermodynamics and corresponding Smarr relation. Then we study the phase transition points and $P - V$ criticality of these black holes.
In extended phase space, the mass appears not only as an internal energy, but as an enthalpy of the thermodynamic system, as below
\begin{equation}
H = M = \frac{1}{3}\pi Pr_0^4 - \frac{1}{6}a{r_0} + \frac{1}{4}kr_0^2 + \frac{1}{2}{k^2}\lambda _{gb}  + \frac{1}{4}{m^2}\left( {c_0^3{c_3}{r_0} + c_0^2{c_2}r_0^2 + {c_0}{c_1}r_0^3/2} \right).
\end{equation}
The Hawking temperature of the  black hole is obtained by applying the definition of surface gravity, is as follows\\
\begin{eqnarray}\label{temp}
\begin{split}
T&=\frac{{{\rm K} ({r_0})}}{{2\pi }}=\frac{1}{{2\pi }}\left[ {\frac{1}{{\sqrt {{g_{rr}}} }}\frac{d}{{dr}}\sqrt { - {g_{tt}}} } \right]{|_{r = {r_0}}} = \frac{1}{{4\pi }}{\partial _r}f(r){|_{r = {r_0}}}\\
&= \frac{{8\pi Pr_0^3 - a + 3k{r_0} + 3{m^2}(c_0^3{c_3}/2 + c_0^2{c_2}{r_0} + 3{c_0}{c_1}r_0^2/4)}}{{6\pi (r_0^2 + 4k\lambda _{gb} )}}.
\end{split}
\end{eqnarray}
The entropy of the black hole is given by using Wald's formula 
\begin{equation}
S = \int_0^{{r_0}} {\frac{1}{T}} \left( {\frac{{\partial M}}{{\partial {r_0}}}} \right)d{r_0} = \pi \left( {\frac{1}{3}r_0^3 + 4k\lambda _{gb} {r_0}} \right),
\end{equation}
which clearly shows the correction of the area law with the GB-term for non-flat topology $(k \ne 0)$. It is also obvious that the entropy of a string cloud and massive term does not affect the black hole entropy.\\
With these definitions, one can obtain the first law of thermodynamics in extended phase space  in the following form
\begin{equation}
dM = TdS + VdP + Ad\lambda _{gb}  + Bda + {C_1}d{c_1} + {C_2}d{c_2} + {C_3}d{c_3},
\end{equation}
with
\begin{equation}
V = {\left( {\frac{{\partial M}}{{\partial P}}} \right)_{S,\lambda _{gb} ,a,{c_i}}} = \frac{1}{3}\pi r_0^4,
\end{equation}
\begin{equation}
A = {\left( {\frac{{\partial M}}{{\partial \lambda _{gb} }}} \right)_{S,P,a,{c_i}}} = \frac{1}{2}{k^2},
\end{equation}
\begin{equation}
B = {\left( {\frac{{\partial M}}{{\partial a}}} \right)_{S,P,\lambda _{gb} ,{c_i}}} =  - \frac{1}{6}{r_0},
\end{equation}
\begin{equation}
{C_1} = {\left( {\frac{{\partial M}}{{\partial {c_1}}}} \right)_{S,P,\lambda _{gb} ,a,{c_2},{c_3}}} = \frac{{{m^2}{c_0}{r_0^3}}}{8},
\end{equation}
\begin{equation}
{C_2} = {\left( {\frac{{\partial M}}{{\partial {c_2}}}} \right)_{S,P,\lambda _{gb} ,a,{c_1},{c_3}}} = \frac{{{m^2}c_0^2r_0^2}}{4},
\end{equation}
\begin{equation}
{C_3} = {\left( {\frac{{\partial M}}{{\partial {c_3}}}} \right)_{S,P,\lambda _{gb} ,a,{c_1},{c_2}}} = \frac{{{m^2}c_0^3{r_0}}}{4},
\end{equation}
where $V$, conjugating thermodynamical variable corresponding to pressure is thermodynamical volume of the black hole and $A,B,{C_i}$'s represent for physical quantities conjugated to the parameters ${\lambda _{gb}},a$ and ${c_i}$'s respectively.
By a dimensional argument one can present the Smarr relation as
\begin{equation}
2M = 3TS - 2VP + 2A{\lambda _{gb}} + Ba - {C_1}{c_1} + {C_3}{c_3}.
\end{equation}
We observe that the massive term ${c_2}$ has scaling weight 0 and it  is constant in the metric function, so, does not appear in Smarr relation. By putting the thermodynamic quantities presented above in Smarr relation or directly from equation (16), we obtain the equation describes the state of the thermodynamic system in the extended phase space, called the equation of  state
\begin{equation}
P = \frac{3}{{4{r_0}}}\left[ {\left( {1 + \frac{{4k{\lambda _{gb}}}}{{r_0^2}}} \right)T - \frac{{3{m^2}{c_0}{c_1}}}{{8\pi }}} \right]  - \frac{{3k + 3{m^2}c_0^2{c_2}}}{{8\pi r_0^2}} + \frac{{2a - 3{m^2}c_0^3{c_3}}}{{16\pi r_0^3}}.
\end{equation}
To investigate critical behavior, one can calculate critical values by using the inflection point properties, ${({{\partial P} \over {\partial {r_0}}})_T} = {({{{\partial ^2}P} \over {\partial {r_0}^2}})_T} = 0$, of the equation of state, as follow
\begin{equation}
{r_c} = \frac{{(2a - 3{m^2}c_0^3{c_3} + 18k{\lambda _{gb}}{m^2}{c_0}{c_1} + \xi )}}{{4(k + {m^2}c_0^2{c_2})}},\nonumber
\end{equation}
\begin{eqnarray}\label{8}\nonumber
{T_c}&=&\frac{1}{{64\vartheta }}\{6a{c_0}{c_1}{m^2}\xi + 16kc_0^2{c_2}{m^2}\xi +16c_0^4{c_2}{m^4}\xi  - 9c_0^4{c_1}{c_3}{m^4}\xi  + 54k{\lambda _{gb}}c_0^2c_1^2{m^4}\xi\\\nonumber
&& + 486k{\lambda _{gb}}c_0^2c_1^2{m^6}+ 288k{\lambda _{gb}}c_0^5{c_1}{c_2}{m^6}- 162k{\lambda _{gb}}c_0^5c_1^2{c_3}{m^6}- 162k{\lambda _{gb}}c_0^5c_1^2{c_3}{m^6} \\\nonumber
&&+486{k^2}\lambda _{gb}^2c_0^3c_1^3{m^6} + 108ak{\lambda _{gb}}c_0^2c_1^2{m^4}+288{k^2}{\lambda _{gb}}c_0^3{c_1}{c_2}{m^4}+  108ak{\lambda _{gb}}c_0^2c_1^2{m^4}\\\nonumber
&& + 288k{\lambda _{gb}}{c_0}{c_1}{m^2}{(k + c_0^2{c_2}{m^2})^2}- 48c_0^7{c_2}{c_3}{m^6}+ 48c_0^7{c_2}{c_3}{m^6} + 27c_0^7{c_1}c_3^2{m^6}\\\nonumber
&&  - 18ac_0^4{c_1}{c_3}{m^4} - 48kc_0^5{c_2}{c_3}{m^4}- 18ac_0^4{c_1}{c_3}{m^4}  + 96kc_0^5{c_2}{c_3}{m^4}\\\nonumber
&&+ 12{a^2}{c_0}{c_1}{m^2}+ 48{k^2}c_0^3{c_3}{m^2}- 32akc_0^2{c_2}{m^2}- 32a{k^2}\\\nonumber
&& + 64kc_0^4{c_2}{m^4}{r_c} + 128{k^2}c_0^2{c_2}{m^2}{r_c}+ 64{k^3}{r_c}\},\\\nonumber
\end{eqnarray}
\begin{eqnarray}\label{9}\nonumber
{P_c} &=& \frac{1}{{32\pi r_c^3}}\{ \frac{{192\pi {T_C}r_c^3(k + c_0^2{c_2}{m^2})\psi }}{{{{(2a - 3c_0^3{c_3}{m^2} + 18k{\lambda _{gb}}{c_0}{c_1}{m^2} + \xi )}^3}}} + 4a\\
&&- 6c_0^3{c_3}{m^2} - 12\pi k{r_c} - 12\pi c_0^2{c_2}{m^2}{r_c} - 9\pi {c_0}{c_1}{m^2}r_c^2\}, 
\end{eqnarray}
where
\begin{eqnarray}\label{11}\nonumber
\xi  &=& (4{a^2} - 12ac_0^3{c_3}{m^2} + 72ak{\lambda _{gb}}{c_0}{c_1}{m^2}\\\nonumber
&&+ 324{k^2}\lambda _{gb}^2c_0^2c_1^2{m^4} - 108k{\lambda _{gb}}c_0^4{c_1}{c_3}{m^4}\\\nonumber
&& + 192k{\lambda _{gb}}{(k + c_0^2{c_2}{m^2})^2} + 9c_0^6c_3^2{m^4}{)^{\frac{1}{2}}},\nonumber
\end{eqnarray}
\begin{eqnarray}\label{10}\nonumber
\vartheta   &=&  2\pi kc_0^2{c_2}{m^2}r_c^2 + \pi c_0^4{c_2}{m^4}r_c^2\\\nonumber
&&+ \pi {k^2}r_c^2 + 24\pi {k^2}{\lambda _{gb}}c_0^2{c_2}{m^2}\\\nonumber
&&+ 12\pi k{\lambda _{gb}}c_0^4{c_2}{m^4}+12\pi {k^3}{\lambda _{gb}},\nonumber
\end{eqnarray}
\begin{eqnarray}\label{12}\nonumber
\psi  &=& 4{a^2} - 12ac_0^3{c_3}{m^2} + 72ak{\lambda _{gb}}{c_0}{c_1}{m^2} + (2a - 3c_0^3{c_3}{m^2}+ 18k{\lambda _{gb}}{c_0}{c_1}{m^2})\xi \\\nonumber
&&+ 9c_0^6c_3^2{m^4} - 108k{\lambda _{gb}}c_0^4{c_1}{c_3}{m^4}+ 162k{\lambda _{gb}}{c_0}{c_1}{m^4} + 162{k^2}\lambda _{gb}^2c_0^2c_1^2{m^4}\\ 
&&+ 96k{\lambda _{gb}}{(k + c_0^2{c_2}{m^2})^2} + 32{k^3}{\lambda _{gb}} +64{k^2}{\lambda _{gb}}c_0^2{c_2}{m^2} + 32k{\lambda _{gb}}c_0^4{c_2}{m^4}. 
\end{eqnarray}
${r_c},{T_c}$ and ${P_c}$ are called horizon radius, temperature and pressure, respectively.\\
The other thermodynamic quantity that can be calculated in the extended phase space for the considered black hole, is the Gibbs free energy given as 
\begin{equation}\nonumber
G = M - TS = \frac{{\Gamma  + 4k{\lambda _{gb}}\Delta }}{{r_0^2 + 4k{\lambda _{gb}}}},
\end{equation}
with
\begin{equation}\nonumber
\Gamma  =  - \frac{\pi }{9}Pr_0^6 + \frac{1}{{12}}(k + c_0^2{c_2}{m^2})r_0^4 - \frac{1}{{18}}(2a - 3c_0^3{c_3}{m^2})r_0^3 + \frac{1}{2}{k^2}{\lambda _{gb}}r_0^2,
\end{equation}
\begin{equation}
\Delta  =  - \pi Pr_0^4 - \frac{{{m^2}}}{4}({c_0}{c_1}r_0^3 + c_0^2{c_2}r_0^2) - \frac{1}{4}kr_0^2 + \frac{1}{2}{k^2}{\lambda _{gb}}.
\end{equation}
In a phase transition, always a thermodynamic potential such as Gibbs free energy becomes non-analytic(discontinuous). To further investigation of critical behavior and effects of GB, massive and cloud string parameters on the criticality, one can plot $P - {r_0}$ and $G - T$ diagrams (Figure 2, 3, 4, 5, 6 and 7) and observe Van der Waals-like behavior clearly. Various diagrams are based on the increase of the GB, massive and cloud string parameters. From $P - {r_0}$ diagrams, it can be seen that by decreasing the value of the GB parameter and increasing massive parameter, Van der Waals-like behavior becomes more apparent. Also, the increase of the value of the cloud string parameter in comparison with the two above-mentioned parameters leads to a diminution of critical behavior. It can be shown that there is no critical behavior and phase transition for values greater than a critical value of the cloud string parameter, ${a_c}$, which satisfies the following condition
 \begin{align}\label{13}
{\left( { - \frac{1}{{27}}{\Im ^3} + \frac{4}{3}\Im \aleph  - \frac{1}{2}{\Re ^2}} \right)^2} - {\left( {\frac{1}{9}{\Im ^2} + \frac{4}{3}\aleph } \right)^3} > 0,\nonumber\\
\Im  = \frac{1}{\Lambda }\left( {3k + 3{m^2}c_0^2{c_2} + 12k{\lambda _{gb}}\Lambda } \right),\nonumber\\
\Re  = \frac{1}{\Lambda }\left( {3{m^2}c_0^3{c_3} - 2{a_c} - 18k{\lambda _{gb}}{m^2}{c_0}{c_1}} \right),\nonumber\\
\aleph  = \frac{1}{\Lambda }\left( { - 12{m^2}c_0^2{c_2}k{\lambda _{gb}} - 12{k^2}{\lambda _{gb}}} \right),
  \end{align}
that is, the above inequality holds for $a \le {a_c}$.\\
In addition, the appearance of discontinuity and swallow tail  in $G - T$ diagrams corroborates critical behavior and the occurrence of phase transition. It is obvious that the incremental effect of the massive parameter on criticality is in contrast to the GB and cloud string parameters. In figure 5, an significant increase in the value of GB parameter has reduced the criticality, which is more evident in flat topology. In figure 6, we see that increasing the massive parameter from zero to a common value causes the emergence of criticality and phase transition. Finally, in figure 7, it can be seen that the increase in the value of the string cloud parameter exceeds the critical value, eliminates the criticality. 
\begin{figure}[ht]
    \centering
\subfloat[a]{\includegraphics[width=4.5cm]{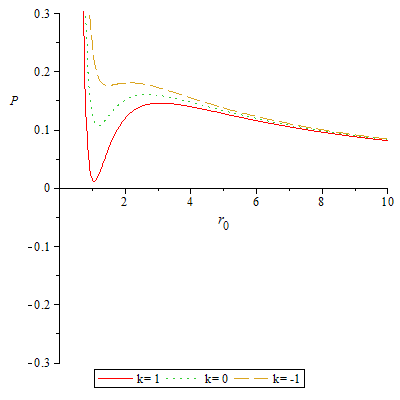}}
\qquad
\subfloat[b]{\includegraphics[width=4.5cm]{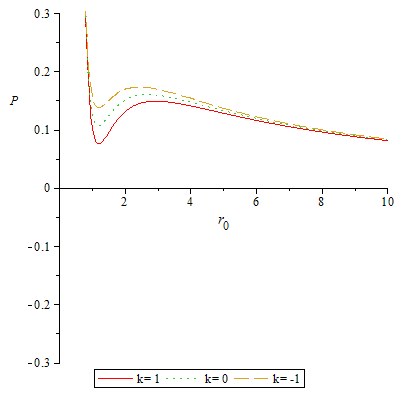}}
\qquad
\subfloat[c]{\includegraphics[width=4.5cm]{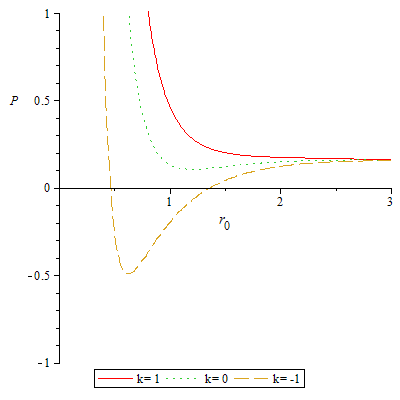}}
\caption{ $P- {r_0}$  for $d=5$, $T = 0.3$, ${c_0} = 1$, ${c_1} = -2$, ${c_2} = 3.75$, ${c_3} = -4$, $a=1$, and $m = 2.1$; (a) :  ${\lambda _{gb}} = 0.0001$, (b): ${\lambda _{gb}} = 0.1$, (c):  ${\lambda _{gb}} = 0.5$.}
\end{figure}
\begin{figure}[ht]
    \centering
\subfloat[a]{\includegraphics[width=4.5cm]{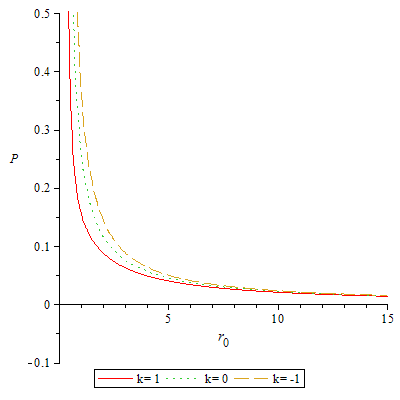}}
\qquad
\subfloat[b]{\includegraphics[width=4.5cm]{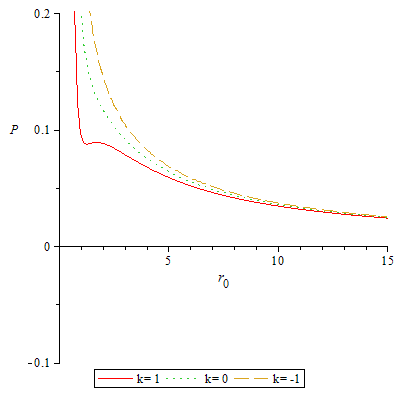}}
\qquad
\subfloat[c]{\includegraphics[width=4.5cm]{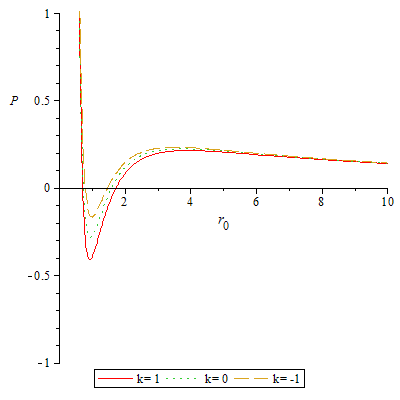}}
\caption{ $P- {r_0}$  for $d=5$, $T = 0.3$, ${c_0} = 1$, ${c_1} = -2$, ${c_2} = 4$, ${c_3} = -4$, $a=1$, and  ${\lambda _{gb}} = 0.01$; (a): $m = 0$, (b): $m = 1$, (c): $m = 3$.}
\end{figure}
\begin{figure}[ht]
    \centering
\subfloat[a]{\includegraphics[width=4.5cm]{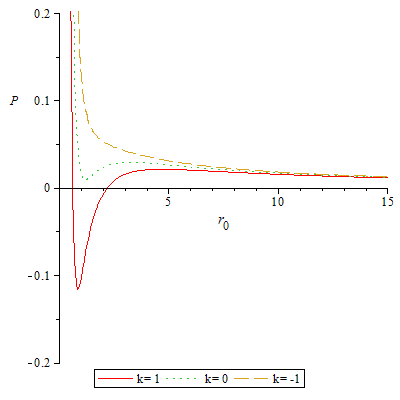}}
\qquad
\subfloat[b]{\includegraphics[width=4.5cm]{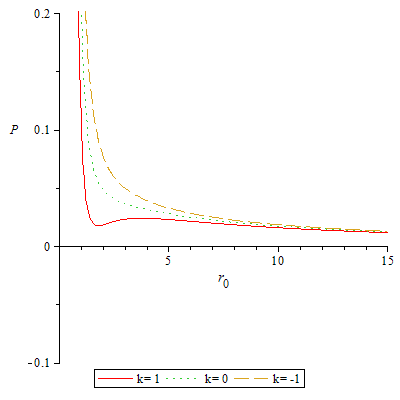}}
\qquad
\subfloat[c]{\includegraphics[width=4.5cm]{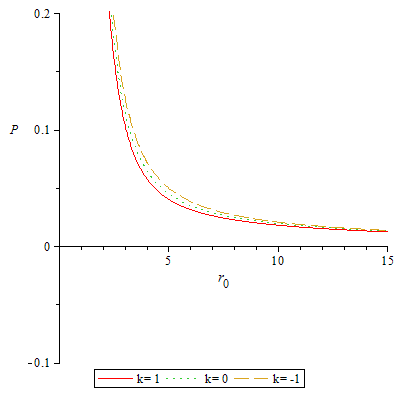}}
\caption{ $P- {r_0}$  for $d=5$, $T = 0.05$, ${c_0} = 1$, ${c_1} = -2$, ${c_2} = 4$, ${c_3} = -4$, ${\lambda _{gb}} = 0.01$, and $m = 1$  ; (a): $a = 1$, (b): $a = 6$, (c): $a = 60$.}
\end{figure}
\begin{figure}[ht]
    \centering
\subfloat[a]{\includegraphics[width=4.5cm]{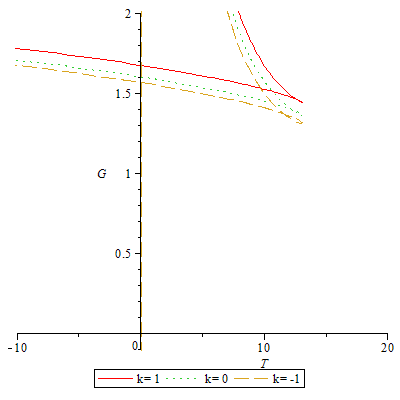}}
\qquad
\subfloat[b]{\includegraphics[width=4.5cm]{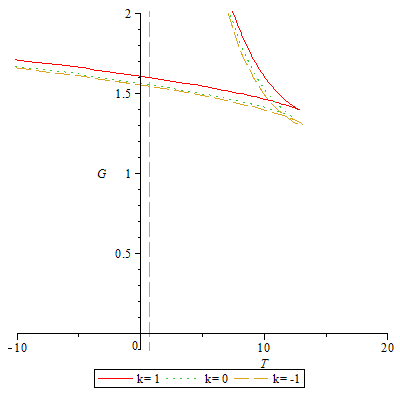}}
\qquad
\subfloat[c]{\includegraphics[width=4.5cm]{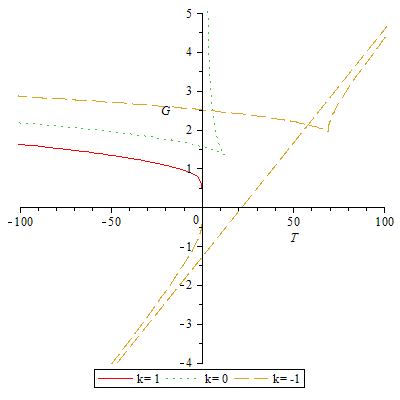}}
\caption{ $G- T$  for $d=5$, $P = 0.33$, ${c_0} = 1$, ${c_1} = -2$, ${c_2} = 4$, ${c_3} = 4$,  $a=1$, and $m = 2.1$; (a) :  ${\lambda _{gb}} = 0.0001$, (b): ${\lambda _{gb}} = 0.01$, (c):  ${\lambda _{gb}} =1$.}
\end{figure}
\begin{figure}[ht]
    \centering
\subfloat[a]{\includegraphics[width=4.5cm]{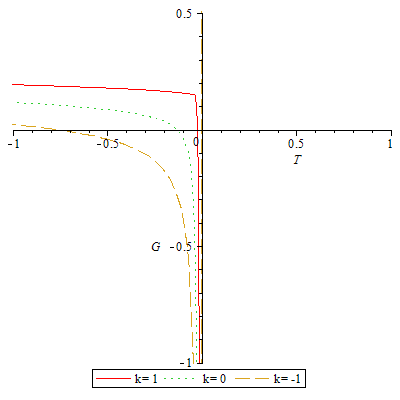}}
\qquad
\subfloat[b]{\includegraphics[width=4.5cm]{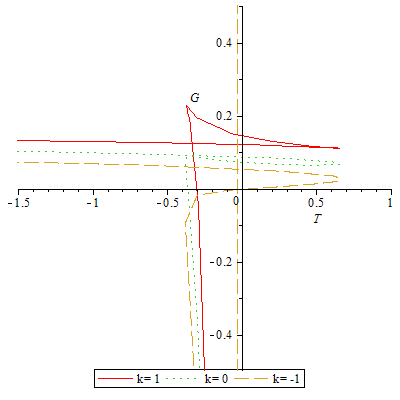}}
\qquad
\subfloat[c]{\includegraphics[width=4.5cm]{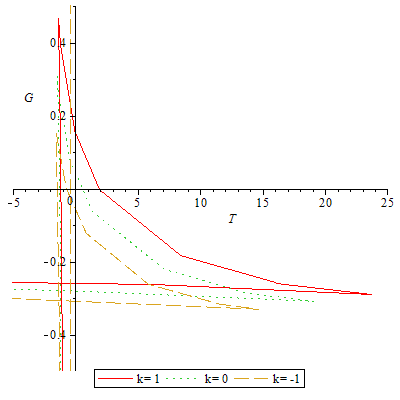}}
\caption{ $G- T$  for $d=5$, $P = 0.033$, ${c_0} = 1$, ${c_1} = -2$, ${c_2} = 4$, ${c_3} = -4$, ${\lambda _{gb}} = 0.0001$, and $a=1$; (a) : $m = 0$, (b): $m = 1$, (c): $m = 2$.}
\end{figure}

\begin{figure}[ht]
    \centering
\subfloat[a]{\includegraphics[width=4.5cm]{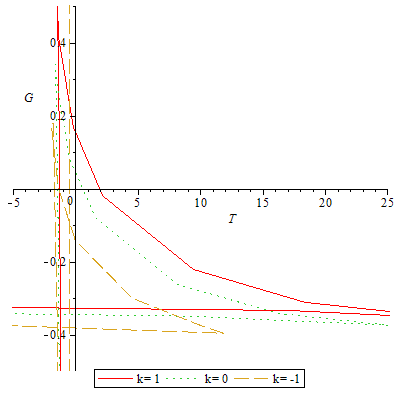}}
\qquad
\subfloat[b]{\includegraphics[width=4.5cm]{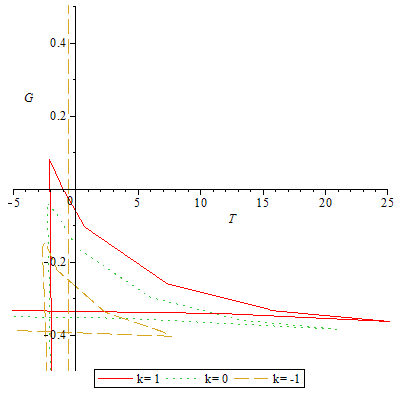}}
\qquad
\subfloat[c]{\includegraphics[width=4.5cm]{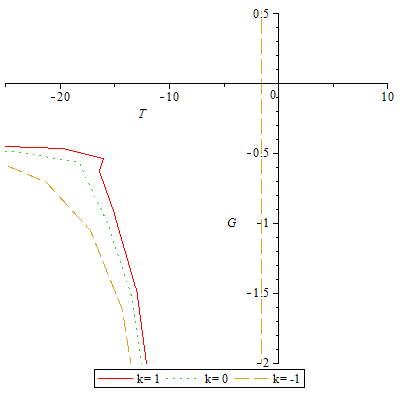}}
\caption{ $G- T$  for $d=5$, $P = 0.033$, ${c_0} = 1$, ${c_1} = -2$, ${c_2} = 4$, ${c_3} = -4$, ${\lambda _{gb}} = 0.0001$, and $m = 2.1$; (a) : $a= 1$, (b): $a= 6$, (c): $a= 60$.}
\end{figure}
So far we have only investigated the critical behavior of the system by fixing temperature. One can study the critical behavior of black hole temperature in $P - {r_0}$ plane and investigate the effect of GB, massive and a cloud of strings parameters on that (Figures 8, 9 and 10). We observe that when the temperature is above the critical value, a behavior similar to that of an ideal gas appears, which is referred to as the ideal gas phase transition. But for temperatures below the critical temperature, three branches are seen, representing small, medium, and large black holes. Except for the latter, which is unstable, the other two are stable and consistent with the Van der Waals liquid/gas phase transition. As we can see in Figure 9, we find that for temperatures below the critical temperature, the pressure decreases with mass reduction being negative. Also, as seen in Figure 10-(a), there exist a particular temperature for which we have ${{\partial P} \over {\partial {r_0}}} = P = 0$, similar to what we have seen in Van der Waals fluid before.\\
\begin{figure}[ht]
    \centering
\subfloat[a]{\includegraphics[width=4.5cm]{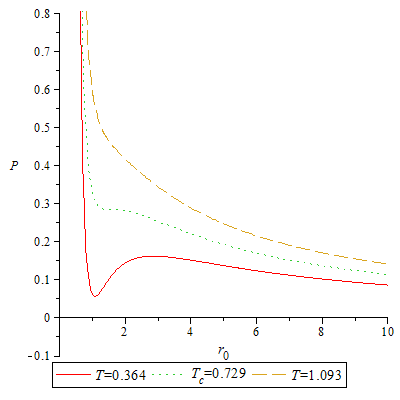}}
\qquad
\subfloat[b]{\includegraphics[width=4.5cm]{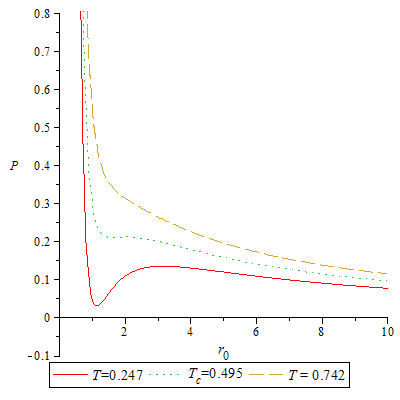}}
\qquad
\subfloat[c]{\includegraphics[width=4.5cm]{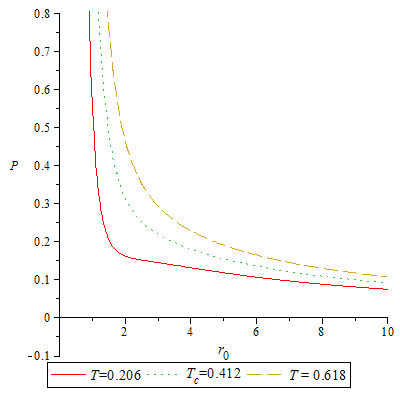}}
\caption{ $P- {r_0}$  for $k=1$, ${c_0} = 1$, ${c_1} = -2$, ${c_2} = 3.75$, ${c_3} = -4$, $a = 1$, and $m = 2.1$  ; (a): ${\lambda _{gb}} = 0.0001$, (b): ${\lambda _{gb}} = 0.1$, (c): ${\lambda _{gb}} = 0.5$.}
\end{figure}
\begin{figure}[ht]
    \centering
\subfloat[a]{\includegraphics[width=4.5cm]{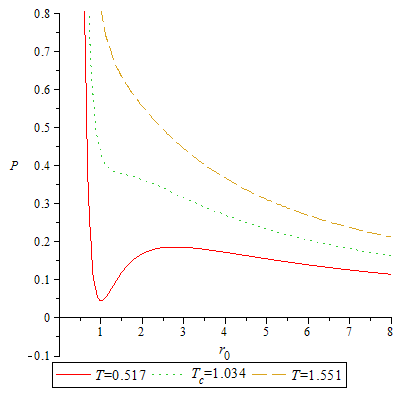}}
\qquad
\subfloat[b]{\includegraphics[width=4.5cm]{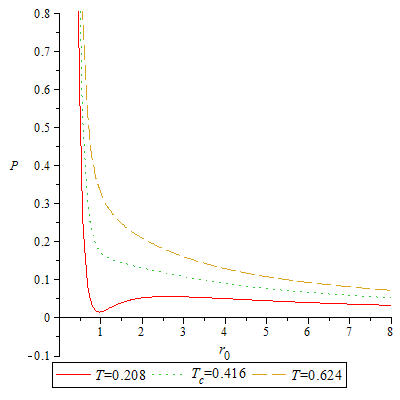}}
\qquad
\subfloat[c]{\includegraphics[width=4.5cm]{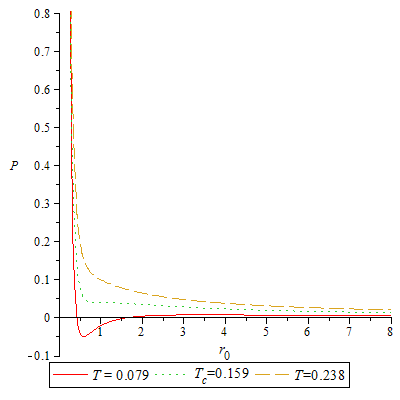}}
\caption{ $P- {r_0}$  for $k=1$, ${c_0} = 1$, ${c_1} = -2$, ${c_2} = 4$, ${c_3} = -4$, ${\lambda _{gb}} = 0.0001$, and $a = 1$  ; (a): $m = 2.1$, (b): $m = 1$, (c): $m = 0$.}
\end{figure}
\begin{figure}[ht]
    \centering
\subfloat[a]{\includegraphics[width=4.5cm]{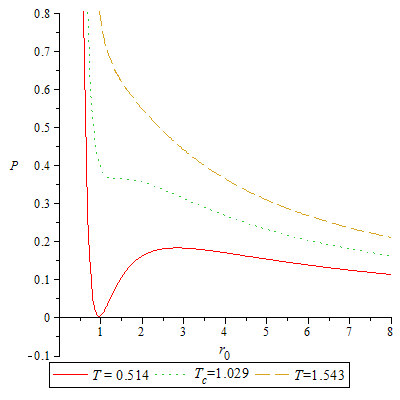}}
\qquad
\subfloat[b]{\includegraphics[width=4.5cm]{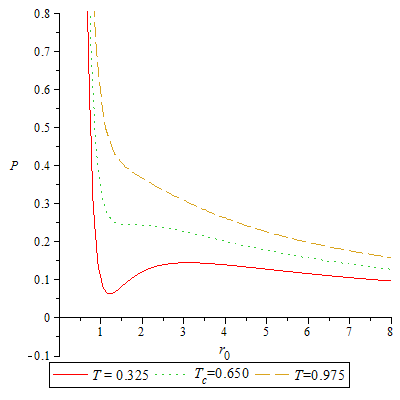}}
\qquad
\subfloat[c]{\includegraphics[width=4.5cm]{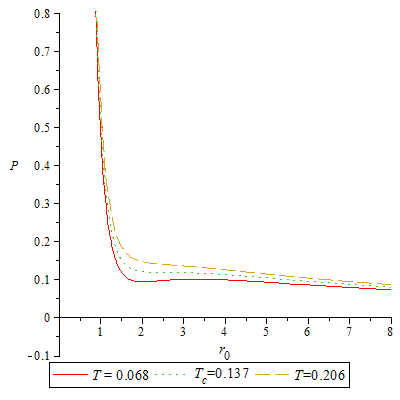}}
\caption{ $P- {r_0}$  for $k=1$, ${c_0} = 1$, ${c_1} = -2$, ${c_2} = 4$, ${c_3} = -4$, ${\lambda _{gb}} = 0.0001$, and $m = 2.1$  ; (a): $a = 0.1$, (b): $a = 6$, (c): $a = 20$.}
\end{figure}
Another way to illustrate the phase transition is to use the $P - T$ diagram for two different phases where the black hole phase transition is between the two so that both phases have the same Gibbs free energy. This phase transition is of the first order and occurs where two surfaces of Gibbs free energy intersect, known as coexistence line in $P - T$ diagrams. At any point on this line, the following equations exist between the two phases mentioned,
\begin{equation}
{G_1} = {G_2}, {T_1} = {T_2}, 2T = {T_1} + {T_2},
\end{equation}
where the indices 1 and 2 correspond to the two different phases of the black hole. The temperature equilibrium of these two phases indicates the isothermal phase transition. We plot equation of pressure with respect to temperature for some values of ${\lambda _{gb}}$, $m$ and $a$ parameters and we  observe the effect of changing these parameters on $P - T$ diagrams  in Figure 11. In this Figure, $p$ is the value of pressure per unit of critical pressure and $\tau $ is the value of temperature per unit of critical temperature.\\
\begin{figure}[ht]
    \centering
\subfloat[a]{\includegraphics[width=4.5cm]{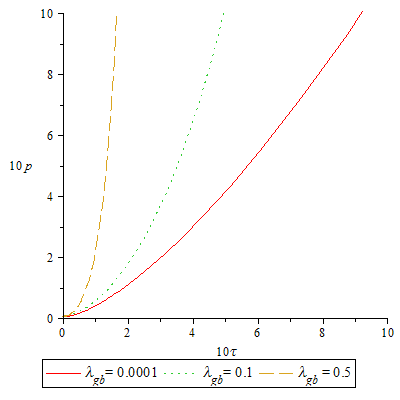}}
\qquad
\subfloat[b]{\includegraphics[width=4.5cm]{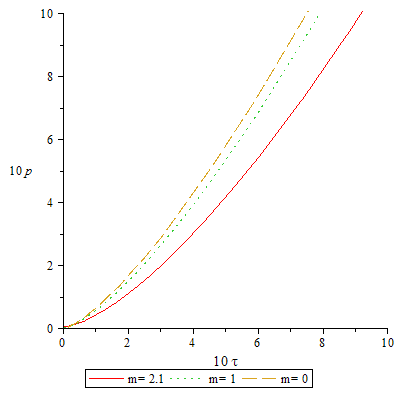}}
\qquad
\subfloat[c]{\includegraphics[width=4.5cm]{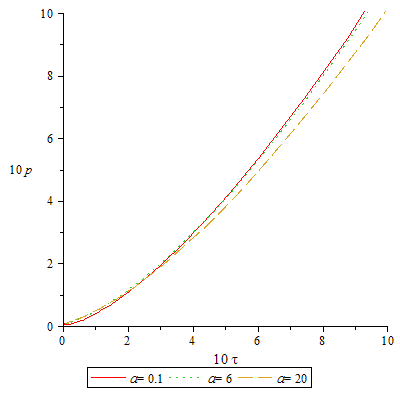}}
\caption{ $P- T$  for $k=1$, ${c_0} = 1$, ${c_1} = -2$, ${c_2} = 3.75$, ${c_3} = -4$ ; (a): $a = 1$, $m = 2.1$ and ${\lambda _{gb}} = \{ 0.0001,0.1,0.5\} $, (b): ${\lambda _{gb}} = 0.0001$, $a = 1$ and $m = \{ 2.1,1,0\}$, (c): ${\lambda _{gb}} = 0.0001$, $m = 2.1$ and $a = \{ 0.1,6,20\} $.}
\end{figure}
In addition, plotting $T- P$ diagrams and examining the behavior of the thermodynamic system would be helpful. This method reveals a process known as Joule-Thomson expansion, which describes the change in system temperature relative to pressure at a constant enthalpy. That means we will have an isenthalpic process that can display heating and cooling phases. To identify the phase of the system, we need to denote the Joule-Thomson coefficient as ${\mu _{JT}} = {({{\partial T} \over {\partial P}})_H}$.
If ${\mu _{JT}} > 0$, it indicates the cooling process in which pressure decreases during the expansion and ${\mu _{JT}} < 0$ shows heating in which pressure increases. To plot the $T-P$ curves and find the process type in our model, we use equations (15) and (16) for different values of constant mass $M$ (Figure 12). As can be seen from the diagrams, our process has only one cooling phase and never enters a heating phase as shown in \cite{Okcu:2016tgt} which follows a heating-cooling process in a Joule-Thomson expansion.\\
\begin{figure}[ht]
    \centering
\subfloat[a]{\includegraphics[width=4.5cm]{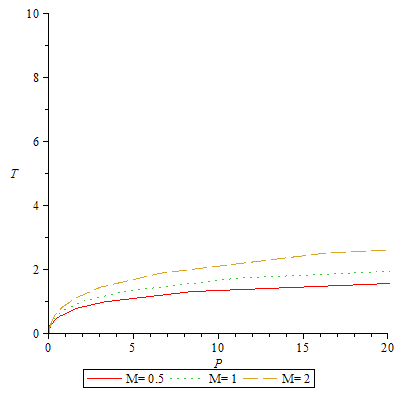}}
\qquad
\subfloat[b]{\includegraphics[width=4.5cm]{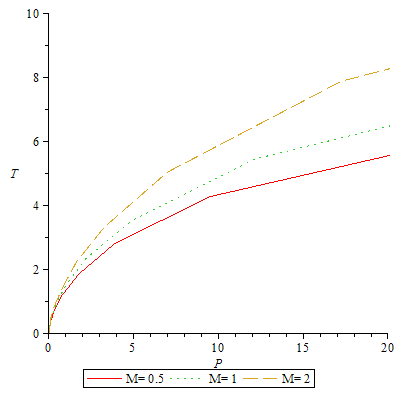}}
\qquad
\subfloat[c]{\includegraphics[width=4.5cm]{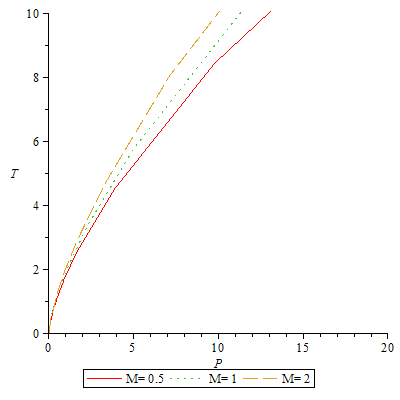}}
\caption{ $T- P$  for $k=1$, ${c_0} = 1$, ${c_1} = -2$, ${c_2} = 3.75$, ${c_3} = -4$, $a = 1$, $m = 2.1$ ; (a): ${\lambda _{gb}} = 0.5$ , (b): ${\lambda _{gb}} = 0.1$ and (c): ${\lambda _{gb}} = 0$. }
\end{figure}
Finally, we study the behavior of physical quantities near the critical point. In order to calculate the critical exponents characterizing the behavior of physical quantities in the vicinity of the critical point, it is advisable to use rescaled quantities $\nu  = {\upsilon  \over {{\upsilon _c}}},\tau  = {T \over {{T_c}}}$ and $ p = {P \over {{P_c}}}$ where $\upsilon  = {4 \over 3}{r_0}$ is specific volume. This simplification makes the equation of state (26) as follows
\begin{equation}
p = \left( {\frac{{{T_c}}}{{{\upsilon _c}{P_c}}}} \right)\frac{\tau }{\nu } + \left( {\frac{{64k{\lambda _{gb}}{T_c}}}{{9\upsilon _c^3{P_c}}}} \right)\frac{\tau }{{{\nu ^3}}} - \left( {\frac{{3{m^2}{c_0}{c_1}}}{{8\pi {\upsilon _c}{P_c}}}} \right)\frac{1}{\nu } - \left( {\frac{{2(k + {m^2}c_0^2{c_2})}}{{3\pi \upsilon _c^2{P_c}}}} \right)\frac{1}{{{\nu ^2}}} + \left( {\frac{{4(2a - 3{m^2}c_0^3{c_3})}}{{27\pi \upsilon _c^3{P_c}}}} \right)\frac{1}{{{\nu ^3}}},
\end{equation}
where is called as law of corresponding state. We can now search for the thermodynamical behavior of the system near the critical points by redefining parameters $t,\omega $ 
\begin{equation}\nonumber
\tau  = 1 + t, \nu  = 1 + \omega.
\end{equation}
That is, the $\nu ,\tau$ and $p$ parameters are expanded around one, so the law of corresponding state would be approximated as
\begin{equation}
p = 1 + \Theta t + \Phi t\omega  + \Omega {\omega ^3} + ...
\end{equation}
where
 \begin{align}\label{14}
&\Theta  = \frac{{9{T_c}\upsilon _c^2 + 64k{\lambda _{gb}}{T_c}}}{{9{P_c}\upsilon _c^3}},\nonumber\\
&\Phi  =  - \frac{{3{T_c}\upsilon _c^2 + 64k{\lambda _{gb}}{T_c}}}{{3{P_c}\upsilon _c^3}},\nonumber\\
&\Omega  = \frac{{72(k + {m^2}c_0^2{c_2}){\upsilon _c} + 27(\frac{3}{8}{m^2}{c_0}{c_1} - \pi {T_c})\upsilon _c^2 - 1920\pi k{\lambda _{gb}}{T_c} - 40(2a - 3{m^2}c_0^3{c_3})}}{{27\pi {P_c}\upsilon _c^3}}.
 \end{align}
To characterize the critical behavior near the critical point, one can introduce the critical exponents as \cite{Lee:2014dha}
 \begin{align}\label{15}
{C_\upsilon } = T\frac{{\partial S}}{{\partial T}}\left| {_\upsilon } \right. \propto {\left| t \right|^{ - \alpha }},\nonumber\\
\eta  = {\upsilon _l} - {\upsilon _s} \propto {\left| t \right|^\beta },\nonumber\\
{\kappa _T} =  - \frac{1}{\upsilon }\frac{{\partial \upsilon }}{{\partial P}}\left| {_T} \right. \propto {\left| t \right|^{ - \gamma }},\nonumber\\
\left| {P - {P_c}} \right| \propto {\left| {\upsilon  - {\upsilon _c}} \right|^\delta }.
 \end{align}
As is clear from the above definitions, the exponents $\alpha ,\beta ,\gamma $, and $\delta $ describe the behavior of specific heat with fixed volume, the order parameter $\eta $, the isothermal compressibility coefficient ${\kappa _T}$, and the critical isotherm, respectively. The subscripts $l$ and $s$ represent the large black hole and the small black hole, respectively, in the phase transition process.\\
The entropy $S$ does not depend on the Hawking temperature $T$, so the specific heat at constant volume ${C_\upsilon }$ is equal to zero, consequently the corresponding critical exponent vanishes ($\alpha  = 0$). To estimate the second exponent $\beta$, one can evaluate ${\upsilon _l}$ and ${\upsilon _s}$ to calculate the order parameter. During the phase transition the pressure of the black hole keeps unchanged. It results that the large black hole pressure equals the small black hole pressure, ${p_l} = {p_s}$ for which
\begin{equation}
1 + \Theta t + \Phi t{\omega _l} + \Omega \omega _l^3 = 1 + \Theta t + \Phi t{\omega _s} + \Omega \omega _s^3.
\end{equation}
On the other hand, from the Maxwell's equal area law, one can further obtain
\begin{equation}
\int_{{\omega _l}}^{{\omega _s}} {\omega \frac{{dp}}{{d\omega }}} d\omega  = 0  \to  \Phi t(\omega _l^2 - \omega _s^2) + \frac{3}{2}\Omega (\omega _l^4 - \omega _s^4) = 0.
\end{equation}
With two  above Eqs, one can get
\begin{equation}
{\omega _l} =  - {\omega _s} = \sqrt {\frac{{ - \Phi t}}{\Omega }}. 
\end{equation}
So the order parameter can be derived as
\begin{equation}
\eta  = {\upsilon _l} - {\upsilon _s} = {\upsilon _c}({\omega _l} - {\omega _s}) = 2{\upsilon _c}{\omega _l} \propto \sqrt { - t}, 
\end{equation}
 where this leads to the conclusion that $\beta  = {1 \over 2}$.\\
 The isothermal compressibility can be estimated as follows
\begin{equation}
{\kappa _T} =  - \frac{1}{{{\upsilon _c}(1 + \omega )}}\frac{{\partial \upsilon }}{{\partial \omega }}\frac{{\partial \omega }}{{\partial P}}\left| {_T} \right. \propto  - \frac{1}{{\frac{{\partial p}}{{\partial \omega }}}}\left| {_{\omega  = 0}} \right. =  - \frac{1}{{\Phi t}}.
\end{equation}
From this one can conclude that $\gamma  = 1$. The critical isotherm is an isotherm process at critical temperature $T = {T_c}$ or $t = 0$. Then we can conclude that $p - 1 = \Omega {\omega ^3}$ that leads to $\delta  = 3$. We see that, the values of critical exponents are independent of GB, massive and cloud string parameters.  The critical exponents in our model are the same as those mentioned in other articles \cite{Lee:2014dha}-\cite{Li:2014ixn}, and all the models reviewed have the same scaling laws.

\section{Heat capacity in canonical ensemble vs in the extended phase space}
\label{sec4}
In this section, we first study the stability of the black hole by using heat capacity in canonical ensemble. Then we investigate criticality in extended phase space by employing heat capacity. When the black hole heat capacity is positive, we say that the black hole is thermally stable. On the other side, an unstable black hole may turn into a stable state, which this transition  is called  phase transition. If we calculate the heat capacity of this black hole, using it, we can determine the  type of phase transition. In this way, the presence of roots and divergence points for the heat capacity will represents the type one and the type two phase transition. When the heat capacity is in the form of a fraction, in order to obtain the roots, we set the numerator to zero and for calculation the divergence points, we set the  denominator to zero.\\ The heat capacity is calculated as follows
\begin{equation}\nonumber
C = \frac{{\partial M}}{{\partial T}} = T\left( {\frac{{\partial S}}{{\partial T}}} \right) = \frac{\Upsilon }{\Psi },
\end{equation}
where
\begin{equation}\nonumber
\Upsilon  = 6{\pi ^2}{(r_0^2 + 4k{\lambda _{gb}})^2}\{  - \Lambda r_0^3 + \frac{{9{c_0}{c_1}{m^2}}}{4}r_0^2 + 3(k + c_0^2{c_2}{m^2}){r_0} - a + \frac{3}{2}c_0^3{c_3}{m^2}\} ,
\end{equation}
\begin{equation}
\Psi  =  - \Lambda r_0^4 - 3( k + c_0^2{c_2}{m^2}+4k{\lambda _{gb}}\Lambda  )r_0^2 + (2a - 3c_0^3{c_3}{m^2} + 18k{\lambda _{gb}}{c_0}{c_1}{m^2}){r_0} + 12k{\lambda _{gb}}(k + c_0^2{c_2}{m^2}).
\end{equation}
It is carefully observed in these equations, since the GB parameter is coupled to the curvature factor, in flat topology, the behavior of heat capacity is independent of GB gravity.
To check the black hole's thermal stability, $T - {r_0}$ and $C - {r_0}$ diagrams are ploted(see Figure 13, 14, 15, 16, 17 and 18). For a region of the event horizon, which temperature is negative, the solution is non-physical, and we have removed it from $T- {r_0}$ diagrams. It is observed that, for spherical($k= 1$) and flat($k= 0$) topologies, temperature have one root and for hyperbolic($k= -1$) have two roots. If the radius of the black hole event horizon is called the size of the black hole, there will be a black hole with minimum size and non-zero entropy at zero temperature. As one can see, increasing the value of the massive parameter leads to the formation of  extrema on the $T - {r_0}$ diagrams, indicating type two phase transition(Figure 13). It is also seen in Figure 14 that the decrease of the GB parameter has a direct effect on the formation of type two phase transition. Finally, the increase of the string cloud parameter above the critical value results in the loss of the type two phase transition(Figure 15).\\
It is observed clearly in Figure 16 that, for spherical topology, with the decrease of the GB parameter, the black hole's instability domain becomes larger. In Figure 17, the $C - {r_0}$ diagrams are ploted in three modes in terms of the massive parameter increase. For small values of massive parameter(Figure 17; a), there is no divergence points, ie, the massive term has a direct effect on the presence of type two phase transition. When the massive parameter increases sufficiently enough(Figure 17; b and c), the heat capacity has divergencies, indicating type two phase transition, ie, the black hole in these points changes the phase between stable and unstable states. As the massive parameter increases, the black hole's instability domain extends. Finally, in Figure 18, we observe that with the increase of the cloud string parameter, the black hole stability domain grows, as long as the string cloud parameter exceeds the critical value, the solutions become completely stable.\\
\begin{figure}[ht]
    \centering
\subfloat[a]{\includegraphics[width=4.5cm]{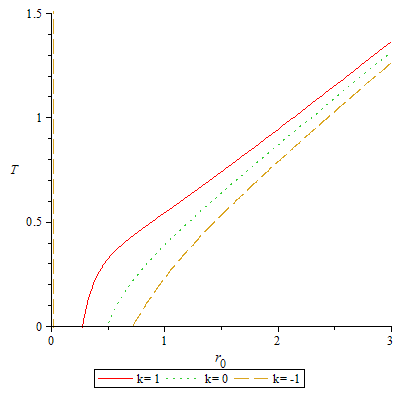}}
\qquad
\subfloat[b]{\includegraphics[width=4.5cm]{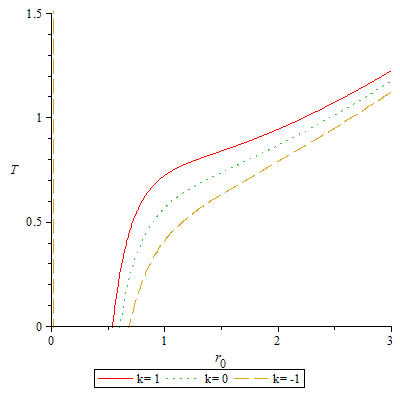}}
\qquad
\subfloat[c]{\includegraphics[width=4.5cm]{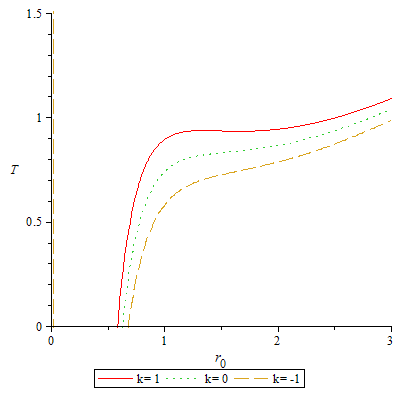}}
\caption{ $T- {r_0}$  for $d=5$, $P = 0.33$, ${c_0} = 1$, ${c_1} = -2$, ${c_2} = 4$, ${c_3} = -4$, ${\lambda _{gb}} = 0.0001$, and $a= 1$; (a) : $m = 0$, (b): $m =1.5$, (c): $m = 2.1$.}
\end{figure}
\begin{figure}[ht]
    \centering
\subfloat[a]{\includegraphics[width=4.5cm]{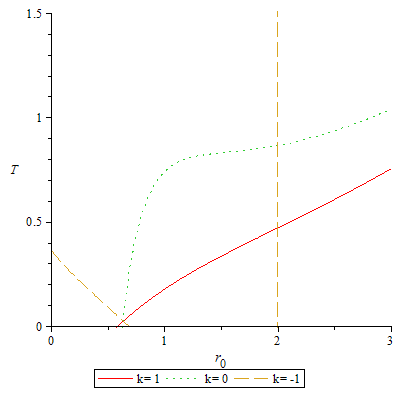}}
\qquad
\subfloat[b]{\includegraphics[width=4.5cm]{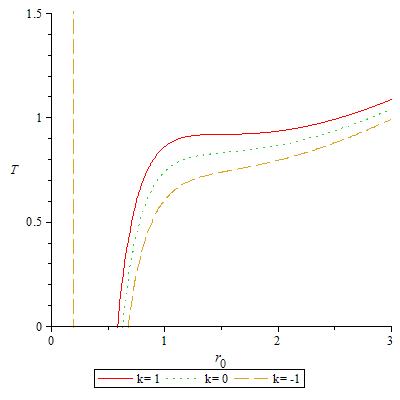}}
\qquad
\subfloat[c]{\includegraphics[width=4.5cm]{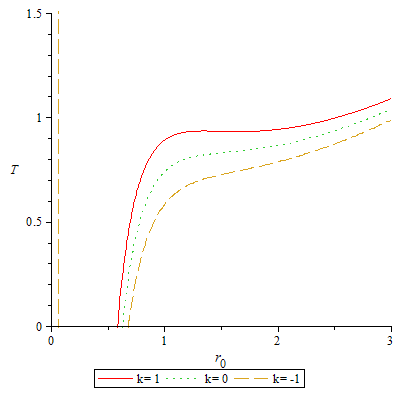}}
\caption{ $T- {r_0}$  for $d=5$, $P = 0.33$, ${c_0} = 1$, ${c_1} = -2$, ${c_2} = 4$, ${c_3} = -4$,  $m = 2.1$, and $a= 1$; (a) : ${\lambda _{gb}} = 1$, (b): ${\lambda _{gb}} = 0.01$, (c): ${\lambda _{gb}} = 0.001$.}
\end{figure}
\begin{figure}[ht]
    \centering
\subfloat[a]{\includegraphics[width=4.5cm]{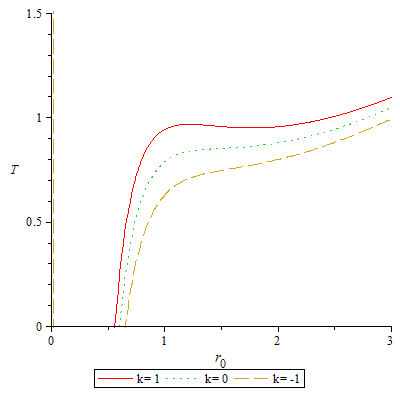}}
\qquad
\subfloat[b]{\includegraphics[width=4.5cm]{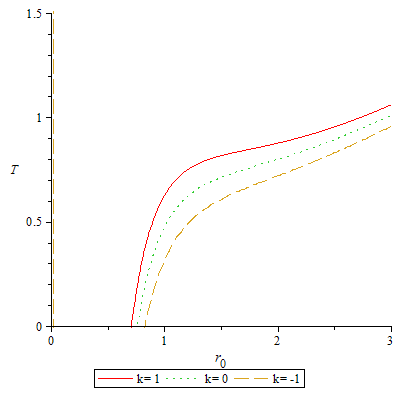}}
\qquad
\subfloat[c]{\includegraphics[width=4.5cm]{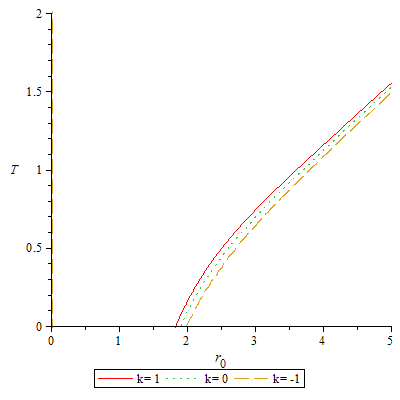}}
\caption{ $T- {r_0}$  for $d=5$, $P = 0.33$, ${c_0} = 1$, ${c_1} = -2$, ${c_2} = 4$, ${c_3} = -4$,  $m = 2.1$, and ${\lambda _{gb}} = 0.0001$; (a) : $a= 0.1$, (b): $a= 6$, (c): $a= 60$.}
\end{figure}
\begin{figure}[ht]
    \centering
\subfloat[a]{\includegraphics[width=4.5cm]{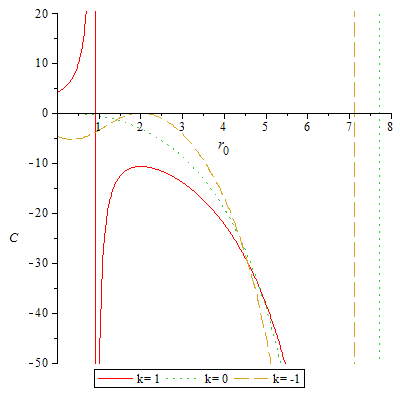}}
\qquad
\subfloat[b]{\includegraphics[width=4.5cm]{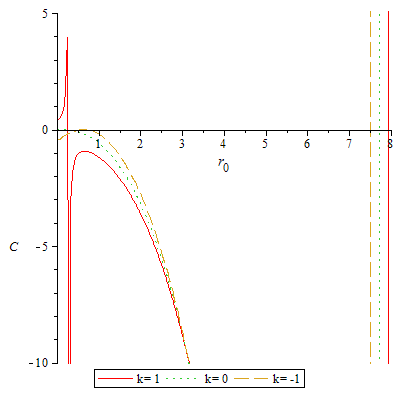}}
\qquad
\subfloat[c]{\includegraphics[width=4.5cm]{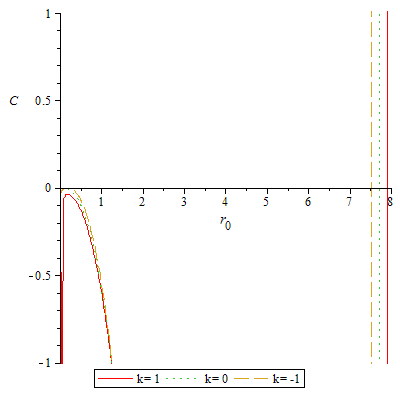}}
\caption{ $C- {r_0}$  for $d=5$, $\Lambda  =  - 1$, ${c_0} = 1$, ${c_1} = -2$, ${c_2} = 4$, ${c_3} = -4$,  $m = 2.1$, and $a= 1$; (a) : ${\lambda _{gb}} = 1$, (b): ${\lambda _{gb}} = 0.1$, (c): ${\lambda _{gb}} = 0.01$.}
\end{figure}
\begin{figure}[ht]
    \centering
\subfloat[a]{\includegraphics[width=4.5cm]{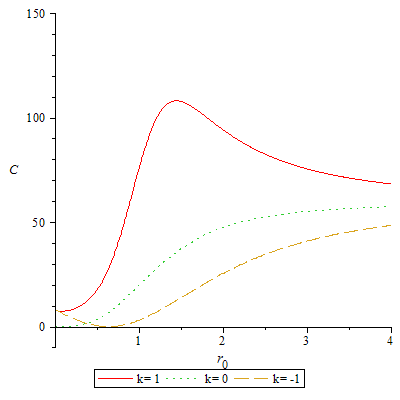}}
\qquad
\subfloat[b]{\includegraphics[width=4.5cm]{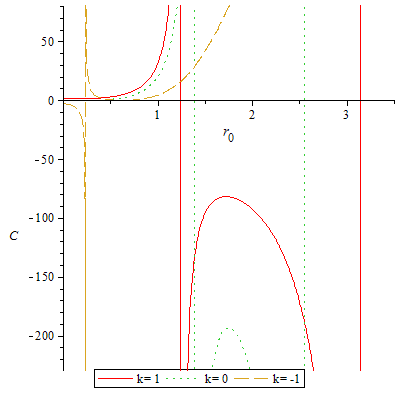}}
\qquad
\subfloat[c]{\includegraphics[width=4.5cm]{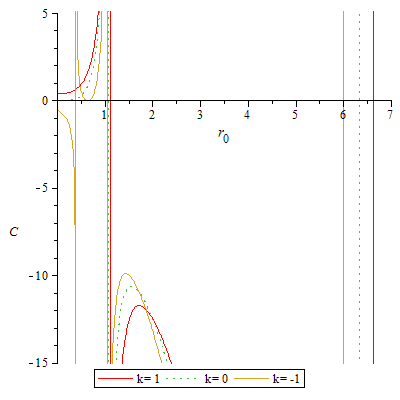}}
\caption{ $C- {r_0}$  for $d=5$, $\Lambda  =  - 1$, ${c_0} = 1$, ${c_1} = -2$, ${c_2} = 4$, ${c_3} = -4$, ${\lambda _{gb}} = 0.1$, and $a= 1$; (a) : $m = 0.05$, (b): $m =1$, (c): $m = 2$.}
\end{figure}
\begin{figure}[ht]
    \centering
\subfloat[a]{\includegraphics[width=4.5cm]{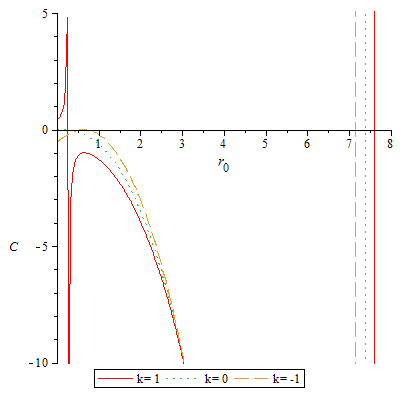}}
\qquad
\subfloat[b]{\includegraphics[width=4.5cm]{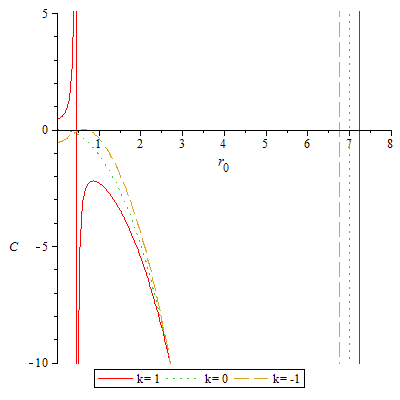}}
\qquad
\subfloat[c]{\includegraphics[width=4.5cm]{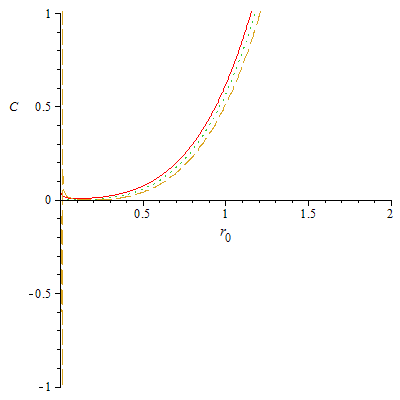}}
\caption{ $C- {r_0}$  for $d=5$, $\Lambda  =  - 1$, ${c_0} = 1$, ${c_1} = -2$, ${c_2} = 4$, ${c_3} = -4$, $m = 2$, and ${\lambda _{gb}} = 0.1$; (a) : $a= 0.1$, (b): $a= 20$, (c): $a= 100$.}
\end{figure}
\begin{figure}[ht]
    \centering
\subfloat[a]{\includegraphics[width=4.5cm]{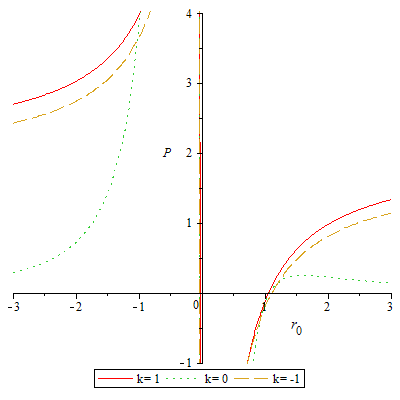}}
\qquad
\subfloat[b]{\includegraphics[width=4.5cm]{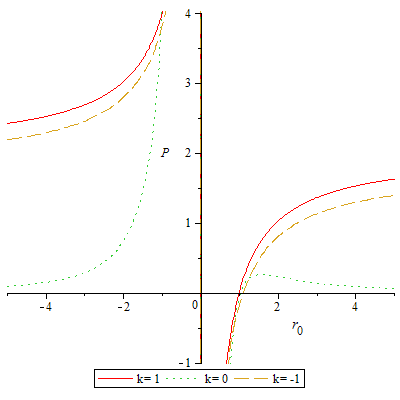}}
\qquad
\subfloat[c]{\includegraphics[width=4.5cm]{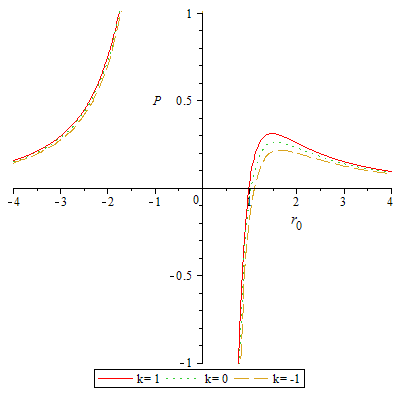}}
\caption{ $P- {r_0}$  for $d=5$, ${c_0} = 1$, ${c_1} = -2$, ${c_2} = 4$, ${c_3} = -4$, $a=1$, and $m = 2$; (a) :  ${\lambda _{gb}} = 0.01$, (b): ${\lambda _{gb}} = 0.0001$, (c):  ${\lambda _{gb}} =0 $.}
\end{figure}
As mentioned earlier, to calculate the divergence points, we must denote the denominator of the fraction of the heat capacity equal to zero. In extended phase space, the critical values in which  phase transition takes place are the same as divergencies in heat capacity. Therefore, as stated in reference \cite{Hendi:2014kha}, by substituting (1) in the denominator of the heat capacity and solving it with respect to pressure, the following relation is obtained
\begin{equation}
P = \frac{{3(k + c_0^2{c_2}{m^2})r_0^2 - (2a - 3c_0^3{c_3}{m^2} + 18k{\lambda _{gb}}{c_0}{c_1}{m^2}){r_0} - 12k{\lambda _{gb}}(k + c_0^2{c_2}{m^2})}}{{8\pi r_0^2(r_0^2 + 12k{\lambda _{gb}})}}
\end{equation}
This equation obtained for pressure is  different from equation (26) calculated for pressure which is called the equation of  state. At the point(s) where this new relation is obtained for pressure, has a maximum(s), phase transition takes place. Indeed, instead of looking for the criticality using the equation of state, we can search for the maximums of the new relation obtained for the pressure from this method. According to the diagrams are plotted in Figure 19, we can say that when ${\lambda _{gb}} \to 0$, the pressure has maximum and the phase transition takes place. Therefore, by using this method, one can calculate critical horizon radius and pressure without referring to the complex formulas given in the usual way.

\indent 
 \section{Conclusion}
\noindent In this paper, we have studied GB-massive black holes in the presence of external string cloud. It is observed that the solutions of the gravitational theory modified by GB term are protected from the causal singularity. It is worth noting that in all of the computed quantities throughout the paper, apart from the metric function, the GB parameter is coupled to the curvature factor, which means that in the case of flat topology $(k = 0)$, GB gravity has no effect. We observe that the first law of the black hole thermodynamics is modified because the black hole entropy gains a contribution from the GB term.\\
We also investigated the criticality and phase transition of the solution in the extended phase space by employing the equation of state. It is important to note that the massive parameter plays a crucial role in creating critical behavior. Also, when ${\lambda _{gb}} \to 0$, the criticality becomes more obvious. It was shown that, the effects of GB and massive parameters on criticality are opposite of each other. Also, the effect of the cloud string parameter on the criticality, like the effect of the GB parameter, is a decreasing effect. In other words, the cloud of strings coupling with the desired gravitational theory has a positive effect on the critical behavior of the theory's solutions, provided that the value of the string cloud parameter does not exceed its critical value. That is, when the cloud string parameter becomes more than critical value, the solutions are completely stable and the phase transition disappears.\\
By studying critical behaviour of the black hole temperature, we conclude that when temperature is above the critical value, Van der Waals-like phase transition would be disappeared and an ideal gas-like behaviour recovers. We also plotted the coexistence line in $P - T$ diagrams in which two phases are in equilibrium and Gibbs free energy and Hawking temperature keep unchanged during transition.\\ 
We investigated $T - P$ diagrams to find out the sign of the Joule-Thomson coefficient. It is shown that the system has only one cooling phase and never enter to a heating phase, that is, the Joule-Thomson effect does not happen. Investigations on the critical exponents showed that the GB, massive and cloud string parameters do not affect them.\\   
Then, we examined thermal stability of these black holes by calculating heat capacity. As expected, it was observed that, the effects of the above three parameters on stability was the opposite of their effects on the criticality.
Finally, using the method mentioned in reference \cite{Hendi:2014kha}, showed that this method is much simpler than the usual method for investigating critical behavior. 



\end{document}